\documentclass{mem}
\usepackage{natbib}\usepackage{txfonts}\usepackage{balance}
\usepackage{graphicx}
\usepackage{txfonts}
\usepackage[a4paper]{hyperref}
\idline{79}{1}
\begin{document}

\def\teff{$T\rm_{eff }$}
\def\kms{$\mathrm {km s}^{-1}$}
\def\logtaustd{$\log\tau_{\rm{5000}}$}

\title{Beyond 1D: spectral line formation 
with 3D hydrodynamical model atmospheres of red giants}


\author{
R. \,Collet\inst{1},
M. \,Asplund\inst{1,2} 
\and R. \,Trampedach\inst{2}
          }


\institute{Max-Planck-Institut f{\"u}r Astrophysik,
Postfach 1317, D--85741 Garching b. M{\"u}nchen, 
Germany; \email{remo@mpa-garching.mpg.de}
\and
Research School of Astronomy and Astrophysics,
Mount Stromlo Observatory,
Cotter Road,
Weston ACT 2611,
Australia}

\authorrunning{Collet et al.}

\titlerunning{3D model atmospheres of red giant stars}

\abstract{We present the results of realistic, 
3D, hydrodynamical, simulations of surface convection 
in red giant stars with varying effective temperatures
and metallicities. We use the convection simulations 
as time-dependent, hydrodynamical, model atmospheres 
to compute spectral line profiles for a number 
of ions and molecules under the assumption of 
local thermodynamic equilibrium (LTE). 
We compare the results with the predictions 
of line formation calculations based on 1D, hydrostatic, 
model stellar atmospheres in order to estimate 
the impact of 3D models on the derivation of 
elemental abundances.
We find large negative 3D$-$1D LTE abundance corrections 
(typically $-0.5$ to $-1$ dex) for weak low-excitation lines 
from molecules and neutral species in the very low 
metallicity cases. 
Finally, we discuss the extent of departures 
from LTE in the case of neutral iron spectral line formation.

}
\maketitle{}

\section{Introduction}
Classical spectroscopic abundance analyses of late-type stars
normally make use of theoretical 1D model atmospheres constructed
under the assumptions of plane-parallel geometry or 
spherical symmetry, hydrostatic equilibrium, and flux constancy.
In late-type stars, the convection zone reaches and affects
the stellar surface layers from which the outgoing stellar flux emerges;
one-dimensional model stellar atmospheres, however, inherently rely
on rather crude implementations of convective energy transport 
such as the mixing length theory \citep{boehm-vitense58} 
or analogous alternative formulations \citep[e.g.][]{canuto91}.
In view of the dynamic and multi-dimensional nature of convection,  
the application of time-independent hydrostatic 1D model atmospheres
to spectroscopic analyses is therefore a potential source of severe 
systematic errors.
Moreover, analyses based on 1D models cannot predict 
strengths and profiles of spectral lines without invoking
ad-hoc fudge parameters (e.g. micro- or macro-turbulence).
In recent years, on the other hand, it has become possible to perform 
realistic 3D hydrodynamical simulations of stellar surface convection
\citep[e.g.][]{nordlund82,nordlund90,stein98,asplund99,freytag02,ludwig02,carlsson04,voegler04}.
In the case of solar-like stars, such
3D simulations have been proven successful in reproducing
self-consistently 
observational constraints
such as the topology of the granulation pattern and the
detailed shape of spectral lines \citep[see review by][]{asplund05}.
Here, we will present some results of
our recent 3D simulations of convection at the surface of
red giant stars \citep{collet06,collet07} focusing in particular 
on their application to spectral line formation.

\section{The hydrodynamical simulations}
\begin{figure*}[]
\resizebox{\hsize}{!}{\includegraphics{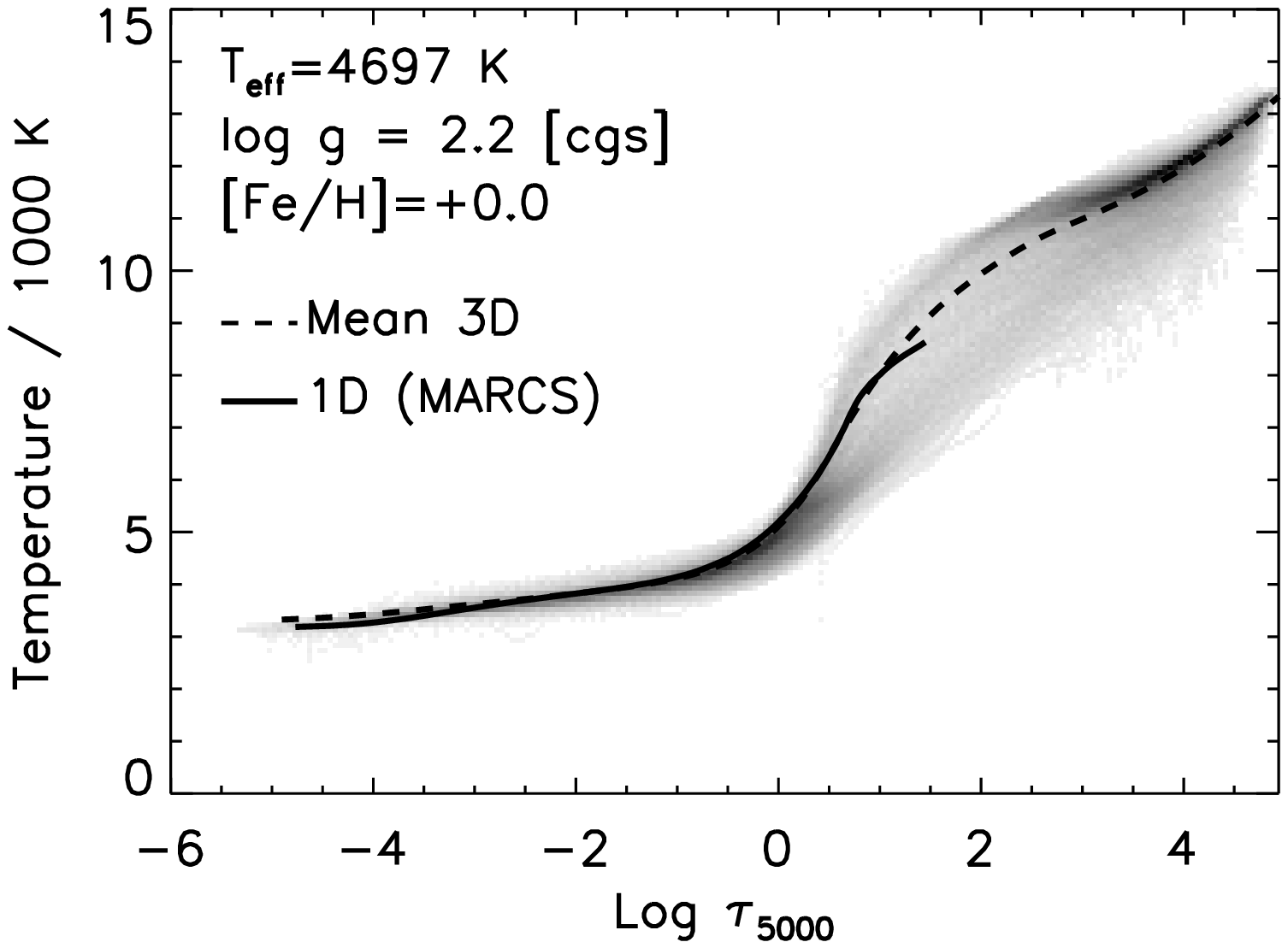}
\includegraphics{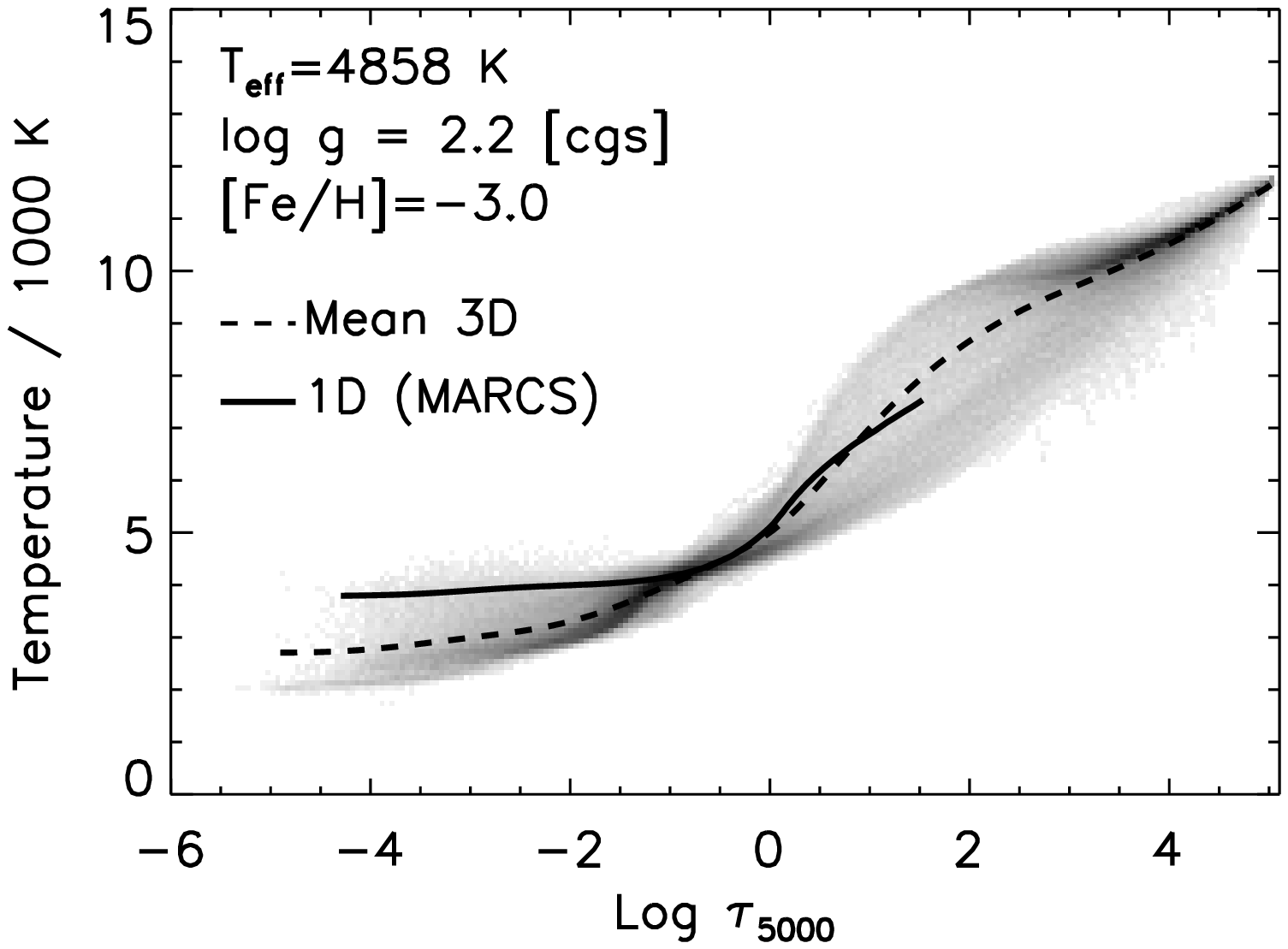}}
\caption{\footnotesize
Atmospheric structures of two 3D red giant simulations
at solar (\emph{left}) and [Fe/H]$-3$ metallicity (\emph{right}).
\emph{Gray shaded area}: temperature distribution as a function of continuum
optical depth at $\lambda=5000$~{\AA}; darker areas indicate values with
higher probability.
\emph{Dashed line}: mean temperature stratification (i.e averaged temperature
on surfaces of constant optical depth) of the 3D model atmospheres.
\emph{Solid line}: temperature stratification of corresponding 1D {\sc marcs} model atmosphere.}
\label{fig:atmos}
\end{figure*}
We use the 3D radiation-hydrodynamics code of
\citet{stein98} to simulate convection at the surface 
of red giants with varying effective temperatures 
(\teff${\approx}4700\,{\ldots}\,5100$~K),
a surface gravity of $\log{g}=2.2$~(cgs), and different
metallicities ([Fe/H]$=0\,{\ldots}-3$).
The radiative-hydrodynamical equations
are solved on a discrete mesh ($100{\times}100{\times}125$)
for a representative volume of stellar surface
covering about eleven pressure scales in depth
and at least ten granules horizontally, and extending
from \logtaustd${\la}-4$ to \logtaustd${\ga}7$ in terms
of continuous optical depth at $\lambda=5000$~\AA.
Open boundaries are employed at the top and bottom of the domain
and periodic boundaries horizontally.
The simulations make use of realistic equation-of-state 
\citep{mihalas88} and opacities 
\citep[][]{gustafsson75,kurucz92,kurucz93}.
We adopt the solar chemical composition from \citet{anders89} with
the abundances of all metals scaled proportionally to [Fe/H].
At each time-step, the radiative transfer equation is solved along the vertical 
and eight inclined rays; 
opacities are grouped into four 
\emph{opacity bins} \citep{nordlund82}
and local thermodynamic equilibrium (LTE) without
scattering terms in the source function
is assumed. 

Qualitatively, the atmospheric structures and gas flows
predicted by the 3D red giant convection simulations
are similar to the ones previously reported for solar-type stars
\citep{stein98,asplund99,asplund01}.
Figure\,\ref{fig:atmos} shows the atmospheric temperature 
structure of two simulation snapshots at solar and [Fe/H]$=-3$ 
metallicity; 
the stratification from 1D, plane-parallel, hydrostatic
{\sc marcs} model atmospheres \citep{gustafsson75,asplund97}
generated for identical stellar parameters, input data, 
and chemical compositions are also displayed.
At metallicities near solar, the mean 3D thermal stratification
of the upper photosphere remains close to the 1D
structure where radiative equilibrium is enforced;
at very low metallicities on the other hand, the surface layers of
3D simulations tend to be substantially cooler than in 1D models.
In 3D, the temperature in the upper photosphere is primarily
regulated by the competition between adiabatic cooling due to the
mechanical expansion of the gas and radiative heating due to
reabsorption by spectral lines of radiation emitted from deeper in. 
At very low metallicity, because of the scarcity and weakness of spectral lines,
adiabatic cooling dominates, hence the balance between cooling and heating 
is reached at lower temperatures than in stationary 1D models 
\citep{asplund99}.

\section{Spectral line formation}

We use the red giant simulations as 3D, hydrodynamical,
time-dependent, model atmospheres to calculate spectral 
line profiles of various ions and molecules under the 
assumption of LTE.
We compute flux profiles for about 80 wavelength points
per spectral line using the same radiative transfer solver 
as in the convection simulations.
We stress that, contrary to 1D calculations, 
3D line formation does not rely upon free parameters such
as micro-or macro-turbulence to account for non-thermal Doppler 
broadening and wavelength shifts due to bulk motions of the gas  
in the photosphere.

We estimate the impact of 3D hydrodynamical models on 
spectroscopic analyses, by differentially comparing 
the \emph{curves-of-growth} of spectral lines computed 
with 3D and 1D {\sc marcs} models using exactly 
the same input physics and numerical scheme.\footnote{For the 1D cases,
we consider two choices of micro-turbulence, $\xi=1.5$~\kms and $\xi=2.0$~\kms.}
Systematic differences between the 3D and 1D thermal structures
as well as velocity gradients and 
temperature and density inhomogeneities present in the 3D models,
can translate into significant differences in terms of
line strengths predicted by the two kinds of models 
for a given chemical composition.
This also implies that the results of
spectroscopic abundance determinations will, in general, depend
on whether 1D or 3D models enter the analysis.

\section{Results}

\begin{figure*}[]
\resizebox{\hsize}{!}{\includegraphics{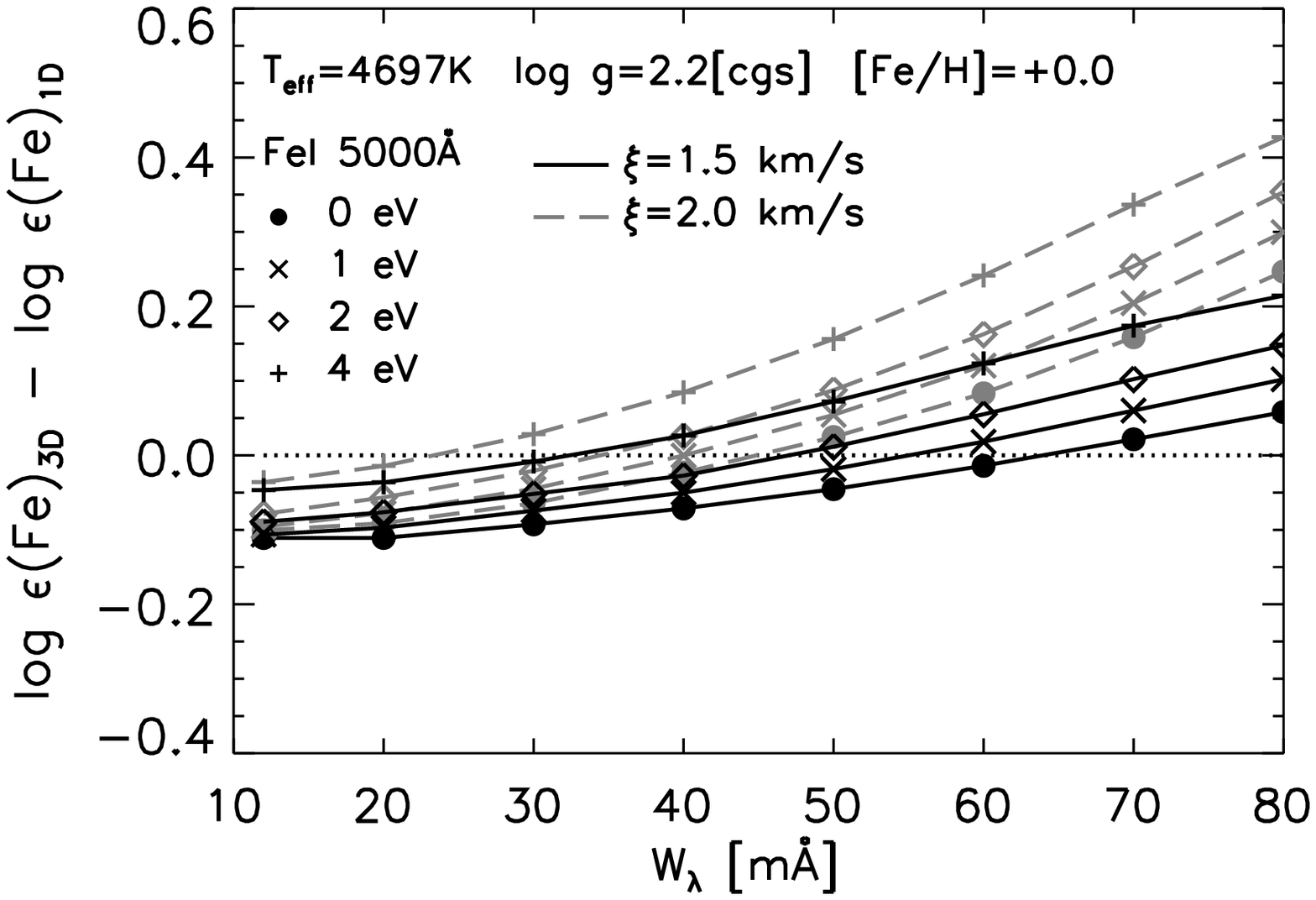}
\includegraphics{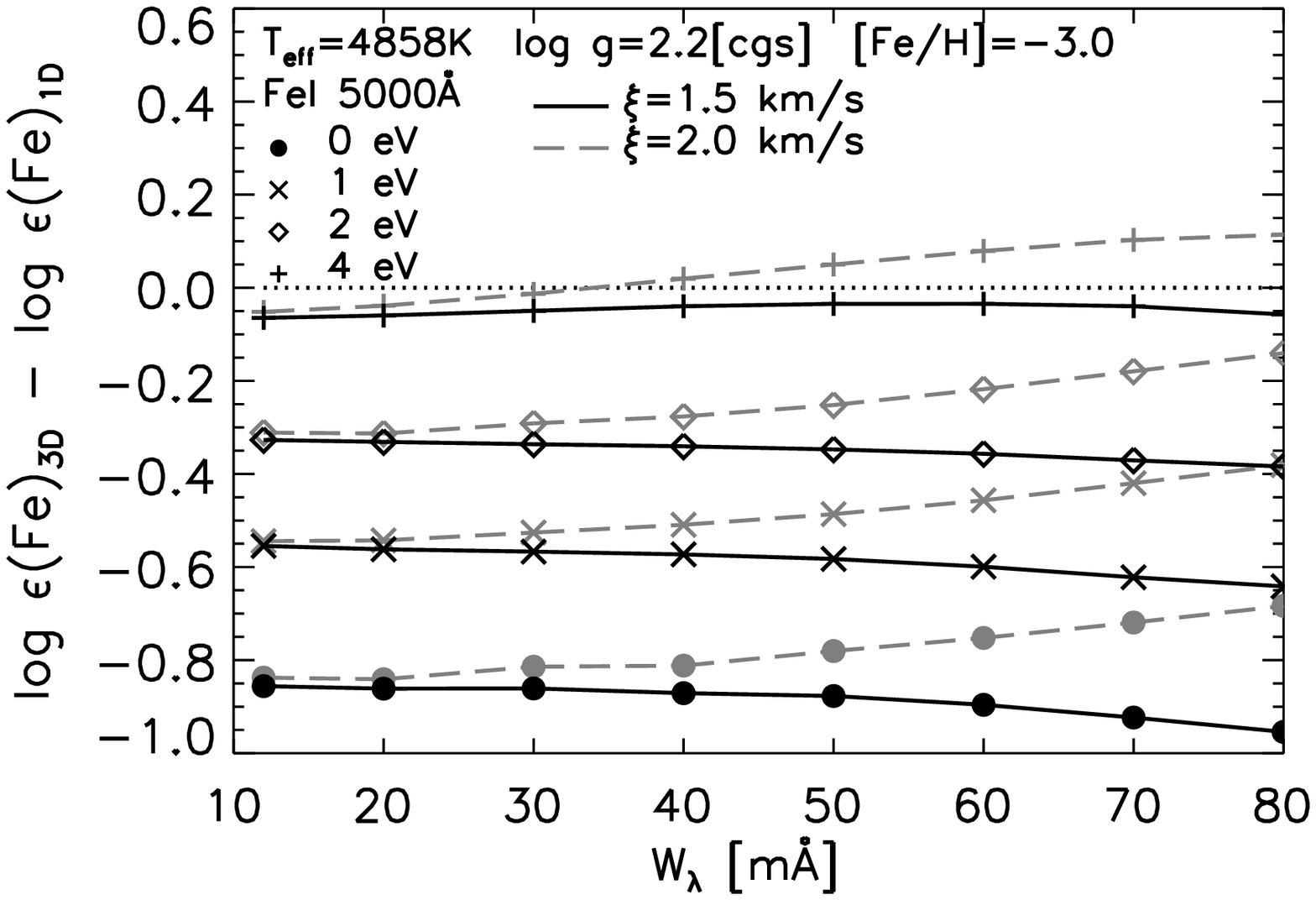}}
\caption{\footnotesize
3D$-$1D LTE Fe abundance corrections derived for 
two red giants at solar (\emph{left panel}) 
and at [Fe/H]$=-3$ metallicity (\emph{right panel})
from \ion{Fe}{i} lines at $\lambda=5000$~{\AA} 
as a function of equivalent width $W_\lambda$.
Corrections are shown as a function of lower level excitation 
potentials of the lines and for two different choices of 
micro-turbulence in the 1D calculations.}
\label{fig:feicorr}
\end{figure*}

\begin{figure*}[]
\resizebox{\hsize}{!}{\includegraphics{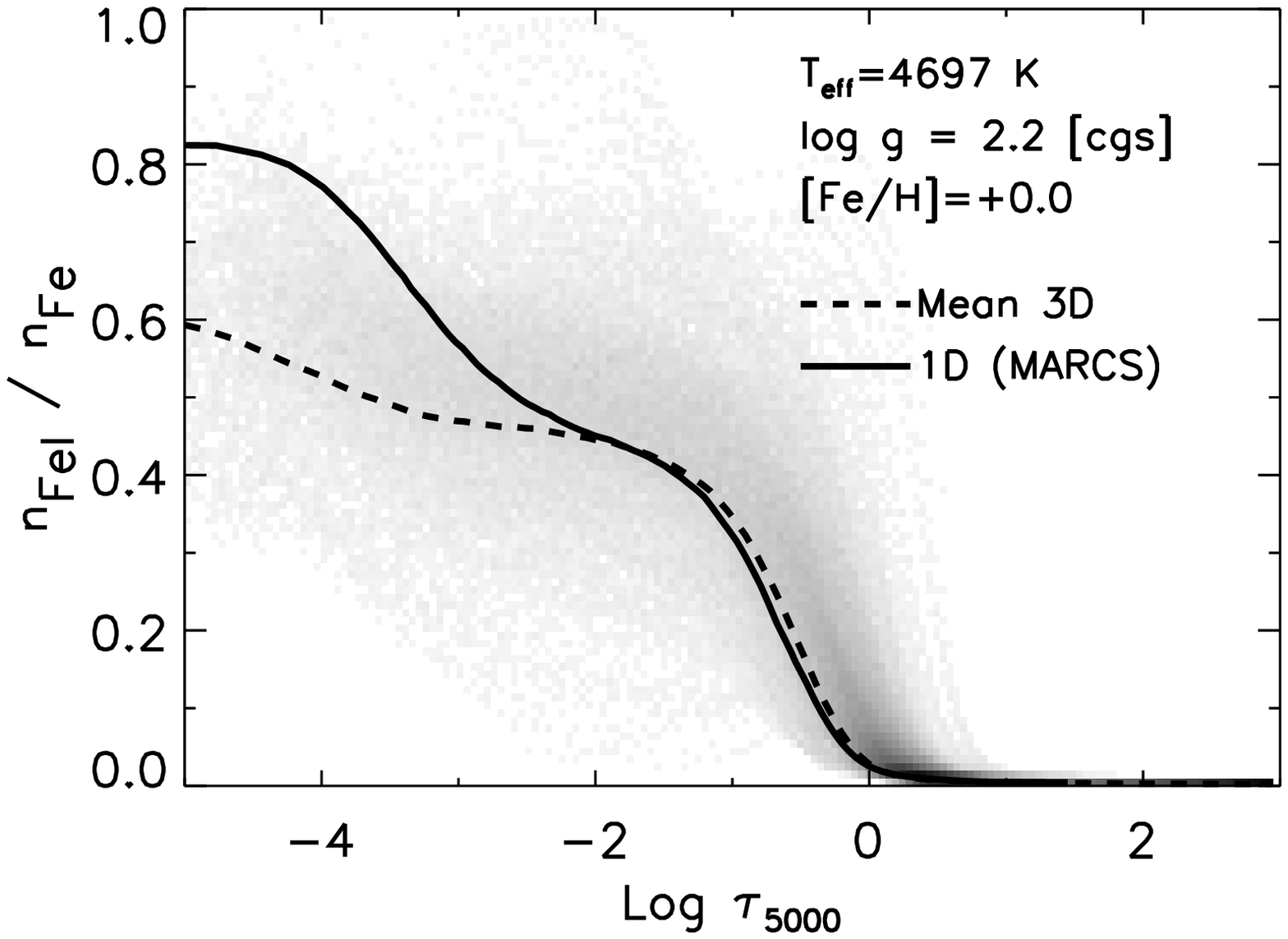}
\includegraphics{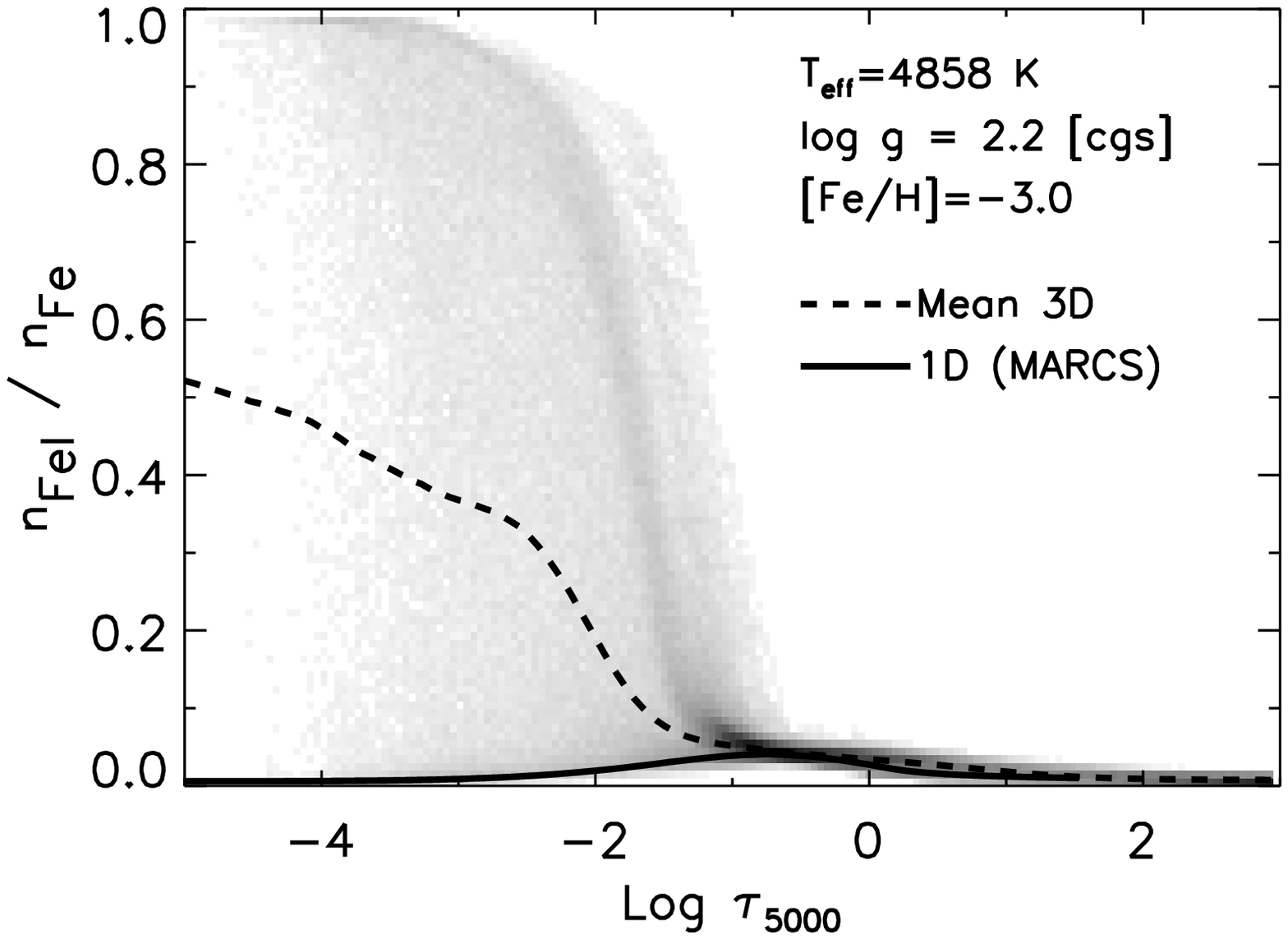}}
\caption{\footnotesize 
Ratio of neutral to total Fe number densities ($n_{\rm{\ion{Fe}{i}}}/n_{\rm{Fe}}$)
as a function of optical depth at $\lambda=5000$~{\AA} 
in the atmosphere of two red giants, at solar (\emph{left panel}) 
and at [Fe/H]$=-3$ metallicity (\emph{right panel}).
The gray shaded area represents the distribution of 
$n_{\rm{\ion{Fe}{i}}}/n_{\rm{Fe}}$ values with optical depth in the 3D model
atmosphere; darker areas indicate values with higher probability. Over-plotted
are the curves for the mean 3D stratifications (\emph{dashed line}) and
the corresponding 1D {\sc marcs} model atmospheres (\emph{solid line}).}
\label{fig:feidens}
\end{figure*}


Figure~\ref{fig:feicorr} shows the differential 
3D$-$1D LTE Fe abundances derived from Fe~{\sc i} lines 
at $\lambda=5000$~{\AA} with varying strengths 
and excitation potentials for two red giants
at solar and very low metallicity.
In LTE, at a given Fe abundance, \emph{weak} Fe~{\sc i} lines 
tend to be \emph{stronger} with 3D models than they do in 1D, 
implying that \emph{negative} 3D$-$1D LTE abundance corrections
are expected. 
Such corrections are more pronounced at low metallicities
and for low-excitation lines:
the differential 3D$-$1D LTE Fe abundances derived from
weak Fe~{\sc i} lines for the model at solar metallicity
are $\la-0.1$~dex, while they can be as large as $-0.8$~dex at
 $\mathrm{[Fe/H]}=-3$.
As line strength increases, differential 3D$-$1D LTE Fe abundances 
remain large and negative in the very metal-poor case,
while they grow positive at solar metallicity.
We note, however, that the trend in the corrections 
for strong lines at solar metallicity might well be an indication
that the present simulations do not resolve the full spectrum
of convective velocities and therefore underestimate 
the non-thermal Doppler broadening. 
Preliminary test simulations of surface convection 
in a red giant at [Fe/H]$=0$ and at a numerical resolution 
of $200{\times}200{\times}250$ show indeed a more prominent 
high-end tail in the distribution of convective velocities 
than at lower resolution.

Qualitatively, we can interpret the behaviour of 
differential 3D$-$1D Fe abundance corrections
by looking at the variations of the number density of neutral
Fe with depth in 1D and 3D model photospheres (Fig.~\ref{fig:feidens}).
At solar metallicity, temperature and density vary with optical
depth in a fairly similar way in both the 1D and \emph{mean} 3D stratifications
and so does the fraction of neutral to total Fe.
Consequently, the 3D$-$1D abundance corrections at solar metallicity
have to be ascribed to temperature and density 
inhomogeneities and velocity gradients in the 3D models 
rather than to differences between the 1D and mean 3D 
structures.
The situation is radically different at very low metallicities.
In the 1D {\sc marcs} stratification, iron is nearly completely
ionized throughout the stellar atmosphere;
in the upper photospheric layers of the 
metal-poor 3D model atmosphere instead, we observe a bimodality
in the distribution of the fraction of neutral Fe
versus optical depth.
While in the hottest parts of the upper 3D photosphere
iron is nearly completely ionized, in the atmospheric regions
where the temperature falls below a critical value (${\sim}3000$~K) 
iron is predominantly neutral.
In other words, due to the systematically cooler temperatures
in the upper photosphere of 3D metal-poor models,  
\ion{Fe}{i} lines tend to be much stronger in 3D than in 1D,
leading to large negative 3D$-$1D LTE corrections
to the Fe abundance.
From a qualitative point of view, 
neutral lines of other elements or molecular features
behave similarly to Fe~{\sc i} lines and are also characterized by
large 3D$-$1D LTE abundance corrections at very low
metallicity. The actual value of the corrections
depend on the details of the ionization and molecular
equilibria.
At this stage, it is important to caution that, \ion{Fe}{i} 
lines are most likely affected by departures from LTE.
The main non-LTE mechanism for \ion{Fe}{i} in late-type stellar atmospheres
is efficient over-ionization driven by the UV radiation field.
This causes an underpopulation of \ion{Fe}{i} levels with respect to LTE,
and leads to \emph{weaker} \ion{Fe}{i} lines, and, in turn,
to the prediction of \emph{higher} Fe abundances in non-LTE.
\citet{collet06} have estimated the non-LTE effects 
on \ion{Fe}{i} lines for the extremely iron-poor giant 
HE\,0107$-$5240 using both a 1D {\sc marcs} model atmosphere 
and a mean atmospheric stratification from 3D simulations 
for the analysis. 
The estimated non-LTE effects (Table\,\ref{tab:feinlte})
are considerable and opposite
to the 3D$-$1D LTE corrections, implying that a combined
treatment of 3D and non-LTE effects is crucial for
accurate Fe abundance determinations. 
\begin{table}
\caption{Fe abundance derived for the red giant 
HE\,0107$-$5240 using a 1D {\sc marcs} model and the mean 
stratification from a 3D model. 
The non-LTE calculations assume the model Fe atom 
from \citet{collet05}, with efficient inelastic H+Fe collisions 
and thermalization of the uppermost\ion{Fe}{i} levels.} 
\label{tab:feinlte}
\begin{center}
\begin{tabular}{ccc}
\hline
\\
Model	&	[Fe/H]$_{\rm{LTE}}$	&	[Fe/H]$_{\rm{non-LTE}}$ \\
\hline
\\
1D       & $-5.40$ & $-4.65$ \\
mean 3D  & $-5.60$ & $-4.75$ \\
\hline

\end{tabular}
\end{center}
\end{table}

\section{Conclusions}
We have presented here some illustrative results of the application
of 3D red giant surface convection simulations to spectral line
formation in LTE.
The differences between the temperature stratifications of 3D
simulations and corresponding 1D model atmospheres
as well as the 3D temperature and density inhomogeneities and 
velocity gradients
can significantly affect the predicted line strengths and, consequently,
the value of elemental abundances inferred from spectral lines.
At very low metallicities, where the deviations of the mean 3D
thermal structure from the classical 1D stratification are largest,
the 3D$-$1D LTE abundance corrections are negative 
and considerable for lines of neutral species 
(about $-0.8$~dex for weak low-excitation \ion{Fe}{i} lines).
Corrections to CNO abundances derived from CH, NH, and OH 
weak low-excitation are also found to be typically in the 
range $-0.5$~dex to $-1.0$~dex in very metal-poor giants
\citep[for further details see][]{collet07}.
Finally, we have also examined possible departures of
\ion{Fe}{i} line formation from LTE; such non-LTE corrections
are opposite to and, according to preliminary 1D test calculations of ours, 
of the same order of magnitude as the ones due to granulation.

\bibliographystyle{aa}
\bibliography{collet}

\end{document}